# User Interests in German Social Science Literature Search – A Large Scale Log Analysis

Daniel Hienert
GESIS – Leibniz Institute for the Social Sciences
Cologne, Germany
daniel.hienert@gesis.org

## ABSTRACT
The social sciences are a broad research field with a lot of sub- and related disciplines. Accordingly, user interests in a digital library for the social sciences are manifold. In this study we analyzed nine years log data of a social science digital library to get an overview of the fields, categories, topics and detailed information needs users are interested in. Based on the log data we have built interactive visualizations which give an overview and concurrently let us look at the detailed interests of users. The underlying log data and the created visualizations are then used to analyze user interests at different hierarchical levels and on a temporal view. The results show that there are topical interests of the users in specific fields and topics of the social sciences but at the same time there exists a diversity of different information needs. Based on these findings we analyze in detail the gap between the indexing language of the system used to annotate documents and the language users apply to articulate their information needs.

## Keywords
Log Analysis; User Interests; Social Sciences; Digital Library

## 1. INTRODUCTION
The nature of the *Social Sciences* to describe the social relationships between humans and their interactions within the society makes them an interdisciplinary field with relations to a lot of other disciplines. The Social Sciences incorporate disciplines such as Sociology, Demography, Ethnology, Political Science, Education, Psychology, Communication Studies, Economics, Social Policy, Interdisciplinary and Applied Social Sciences and related areas such as the Humanities, Legal and Administrative Science, Natural Sciences, Technology and Medicine[1]. Maurice B. Line summarized the characteristics of the social sciences as a challenging discipline for bibliographic control pointing out that there is, among other points, an inherent instability of the subject matter and a lack of terminology that was common over time and across countries [18].

This complexity is also mirrored in digital libraries which collect and offer bibliographic (and other) information for the social sciences. On the level of documents and databases, document collections from different sub fields of the social sciences need to be collected from different origin databases or as a part of larger multidisciplinary databases to get a complete view of the whole field. On the level of semantic integration, knowledge organization systems try to solve the challenges of the complex taxonomy and the multiple database problem with their different thesauri and classification systems. On the human side, there is the challenge that users with a diversity of information needs in the different fields of the social sciences are looking for information. Understanding these information needs and classifying them in a more abstract classification system is a first step to understand the topical interests of users and to build better supporting services.

To observe and capture the complexity of the social sciences on the base of user-system interaction we concentrate in this work on the following two research questions: (1) "*What are the user interests in the (German) Social Sciences on different topical levels and over a long period of time?*" and (2) "*Are knowledge organization systems such as thesauri good enough to capture these user interests?*"

To this purpose we analyze nine years log data from Sowiport, a digital library for the Social Sciences. The goal is to capture the user interests at different levels of main and sub- categories, topics and user information needs. Based on the log data we created different interactive visualizations which allow analyzing the topical interests from overview to detail and on a temporal basis. We will especially investigate the differences between the controlled language used for the indexation of documents and the language used by the users to address their information needs.

## 2. RELATED WORK
In this section we first report on the characteristics of log analysis in general. Then we will show which insights have been captured with log analysis in different domains. The last sub section in particular refers to the information seeking behavior of social scientists.

### 2.1 Log Analysis in General
Log analysis is a method to analyze the interaction between the user and the search system. Kelly [15] arranges log analysis in the middle of a research continuum, one step towards the system`s focus. It can be a compromise between the system and the human focus as user interactions are captured in logs, but the method is scalable and can identify usage patterns for a mass of end users. Peters [25] contributes a literature review about the history and development of Transaction Log Analysis (TLA) until the mid-

---

[1] see for example the definition of the UNESCO at http://portal.unesco.org/en/ev.php-URL_ID=13135&URL_DO=DO_TOPIC&URL_SECTION=201.html in section "Group I - Social sciences, including" or the browsable Universal Decimal Classification (UDC) for the field of the Social Sciences at http://udcdata.info/18549

90s. The article shows the diversity of applications for which TLA can be used: from computing basic statistical data such as the average session length or average keywords entered per user to the field of analyzing more complex user behavior, e.g. to find out what makes a session successful. Jansen [11] gives a good overview of the possibilities and limitations of log analysis and describes that it can identify trends and typical interactions, but cannot record the user's perception of the task, the underlying information need or the underlying situation and context of the search. Agosti et al. [1] explicitly distinguish between web search engine log analysis (WSE) and digital library log analysis (DLS). In WSE users query a search engine with diverse information needs and the retrieved documents are webpages (or other types such as images, videos etc.). In DLS users are searching in a specialized domain or around a specific topic. Here, retrieved documents are (mostly) bibliographic records. These documents are structured and organized in a knowledge organization system (KOS) such as thesauri and classification systems. A second aspect is that domain or topic knowledge can influence search behavior. Wildemuth [29], for example, showed that search tactics of medical students changed over time when the domain knowledge changed. Kelly and Cool [16] show the influence of topic familiarity on searching efficacy and reading time. White et al. [28] conducted a large-scale analysis of web search logs and showed that domain experts had different search behavior than non-experts.

## 2.2 Log Analysis in Different Domains

Log analysis has been conducted in different domains to derive knowledge about the specific information seeking behavior of users in different scientific communities.

Islamaj Dogan et al. [10] for example analyzed the search behavior of *biomedical* users in PubMed by the investigation of one month log data. Among other things they annotated a subset of 10,000 randomly chosen search queries with semantic categories and found that searches by author name, gene/protein and by disease were most frequent. Natarajan et al. [23] conducted a small scale log analysis of an electronic health record system. Here, queries were manually classified to different semantic categories such as "Laboratory or Test Result" or "Disease or Syndrome". Nicholas et al. [24] addressed the issue of integrating log files from five different sources of health information. They present a method to categorize web page titles to health categories on a large scale. For the domain of *chemistry* Davis et al. [2,3] examined transaction logs of electronic journal downloads of 29 journals from the American Chemical Society (ACS) servers. Their studies are based on the prior observation that chemists rely heavily on journal literature. First, they found that the majority of users downloaded only few articles and consulted only few journals [3]. To answer the question how scientists locate published articles in ACS they found that most people came from the electronic library catalog and from bibliographic databases [2]. Ke et al. [14] report results and findings from a transaction log analysis in the domain of *Science, Technology and Medicine*. This work investigates large logs from Elsevier's ScienceDirect website in Taiwan offering bibliographic information and full-text articles from 1,300 journals and about 625,000 users. The authors describe different usage patterns, for example, that full-text article viewing is the most active category; about different user actions (34% browse vs. 13% search) or about the average length of search queries with about 2.27 terms. They also report on common query terms ("acid", "review", "cell" etc.), but did not further process them. Jones et al. [13] contribute a study with a transaction log analysis of user activity in the *Computer Science* Technical Reports Collection of the New Zealand Digital Library. This collection includes about 46,000 technical reports and the authors analyzed a log of 61 weeks with about 30,000 queries. In summary, they found very short sessions, few queries per user, simple queries and small changes for reformulating and, that only the first results were viewed. The authors also present a list with commonly appearing query terms ("compression", "object", "software" etc.) and state that this data could serve as a basis to identify user groups whose information needs have been met or not. Another work in the field of *Computer Science* [20] analyzed six months of usage of the Research Index digital library and found that users preferred relatively brief queries and had short session times. Yi et al. [30] conducted a log analysis of two fields under the umbrella of the social sciences: *Psychology* and *History*. Besides basic query statistics they conducted a more sophisticated analysis of query terms and their ability to categorize queries to subjects. They tested with (1) single query terms, (2) co-occurring word pairs and (3) multiword terms and found that the last method worked best for that purpose. Within the analysis of a multi-disciplinary OPAC (including the field "social sciences and law") Villen-Rueda et al. [27] studied and compared the distribution of queries with author, title or subject search. In their analysis they found low values for subject searches of about 14% in contrast to related literature which report values between 26% and 83%.

Apart from [30] we could not find any transaction log analysis which use specifically logs from a social science digital library. Most work in TLA concentrates very much on statistical measures of session lengths, average query terms etc. In this work we rather want to focus on the topical interests of users on different levels of abstraction.

## 2.3 Information Seeking Behavior of Social Scientists

A study in 1993 by Ellis et al. [5] identified six basic information seeking strategies of social scientists: (1) *starting* (initial search for information), (2) *chaining* (following referential connections such as references), (3) *browsing* (e.g. by scanning of journals or tables of contents), (4) *differentiating* (between sources to filter materials), (5) *monitoring* (the field of interest by following sources such as journals, conferences) and (6) *extracting* (relevant material from sources). In 2006, after the long-time establishment of digital libraries and web-based information retrieval systems, Meho & Tibbo [22] mainly confirmed Ellis' model, but added some new behavior features such as (7) *accessing* (the material found with methods above), (8) *networking* (with colleagues to exchange information), (9) *verifying* (the accuracy of the information), and (10) *information managing* (by archiving and organizing the information).

All lot of these activities are nowadays (at least in a basic manner) supported by modern digital libraries. Users can search for information, follow references and citations, browse in journals or proceedings, see citation counts, access the full text, store information found and make alerts for new search results. This gives some evidence that user`s interests in this field can be concluded from user interaction within digital libraries with large populations.

## 3. A DIGITAL LIBRARY FOR SOCIAL SCIENCE LITERATURE INFORMATION

Sowiport[1] is a digital library for social science information [9] and contains more than nine million bibliographic records, full texts and research projects from twenty-two different databases with German-language and English-language content. The portal reaches about 25,000 unique visitors per week, mainly from German speaking countries.

The bibliographic records contained in Sowiport are mostly annotated with keywords and sometimes with category information depending on the source database. For example, for records of the most requested German database SOLIS (Social Science Literature Information System) both information are available. Here, 475,000 documents are manually annotated with keywords from the thesaurus for the social sciences (TheSoz[2]) with about 12,000 entries and categorized with the classification for the social sciences[3] with 14 main classes and 145 sub classes. However, other databases included in Sowiport use different thesauri and classification systems what makes it difficult to deploy an overall indexing and classification system. To solve this issue we have created the heterogeneity service (HTS) which expands the user's search query with equivalents from other thesauri. The HTS contains cross-concordances [21] for 25 different thesauri with about 513,000 controlled terms. When a search query is submitted by a user of Sowiport it will be internally expanded with equivalent terms from the HTS. For example, the query term "workforce" is expanded with terms like "labor force", "manpower", "employees" and "workers".

A second service is the Search Term Recommender (STR) [19] which can map arbitrary user terms to controlled terms from a subject-specific vocabulary. The system is based on a co-occurrence analysis with free terms from titles and abstracts and controlled terms from descriptors. A logarithmically modified Jaccard similarity measure is used to rank term suggestions for a given input term. The STR service is included in the term recommender of Sowiport to propose controlled terms for free user inputs. If, for example, the user enters the term "hunger" into the search bar, the system proposes terms like "nutrition", "food", "poverty", "Third World" and "developing country".

## 4. LOG DATA PROCESSING

In this section we describe the different preprocessing steps from raw log data files to explicit user actions with keyword and category information.

### 4.1 Building User Actions

The transaction logs for the analysis of user interest in the social sciences were generated by two different technical systems underlying Sowiport. From 2007 until April 2014 Sowiport based on an own-developed JAVA framework and a SOLR search engine (sowiport-old). Since 2014 the digital library framework VuFind (sowiport-new) with a SOLR search engine is used.

For both frameworks all search actions were logged in a database table. This table was expanded for each search query with the first ten search results as a list of document IDs. These IDs have been either recorded in separate Tomcat log files and were matched in a preprocessing step by the session-id and the search query (for sowiport-old) or were directly logged in the database table (sowiport-new). Additionally, all search engine robots in the table were marked with the help of a heuristic. All rows with the same IP and with more than 100 rows in the table were compared to a number of patterns which indicate a link call from URLs that had been propagated in search engines. If only these patterns were matched and no other actions were found the IP was marked as a bot and excluded from further processing. For sowiport-new bots have been additionally excluded from logging if the request header has been given any evidence for a bot.

We only used *initial searches* from users which (1) entered new or add additional terms in the simple or advanced search form and submitted the search or (2) initiate a new search by clicking on authors, keywords, classifications, journals or proceedings. We did not take search modifications such as paging, sorting, filtering, faceting, exporting or any other adaption into account. This way, search topics are not count multiple times per user.

In a next step we extracted two user actions from these tables: (1) a user viewing the detailed information for a document record ("*view_record*") and (2) a search query conducted by a user ("*search*"). Both actions were identified with different patterns for sowiport-old and sowiport-new and the actions with session-id, date, action type, user search terms, viewed document-ids or search results list ids were copied to a new table. The final table contains 8,799,694 user actions from November 2007 to July 2016 with 2,148,414 searches and 6,651,280 record views.

### 4.2 An Overall Keyword & Classification System

In the next processing step we want to apply keyword and category information to each user action. The information therefore derives from the record which has been viewed (for the *view_record* action) or from the results of the search query (for the *search* action). In the previous section we already addressed the issue of different thesauri and classification systems of the twenty-two different source databases included in Sowiport. Consequently, documents appearing in our user actions are annotated with different thesauri and classification information. To achieve a canonical categorization we decided to use the thesaurus for the social sciences TheSoz and the classification for the social sciences (explained in Section 3) as an overall classification system. This has some benefits: (1) a large part of the viewed documents and search results contain SOLIS documents (the log data shows that more than half of user actions contain SOLIS records), (2) TheSoz descriptors and classifications are manually added to SOLIS documents and the system is of a high quality and has been improved over the years by information professionals, (3) we have already developed a set of services such as HTS and STR (see Section 3) with which we can transform keywords from other thesauri and from uncontrolled terms to descriptors of the thesaurus for the social sciences.

For user actions which already included one or more SOLIS documents (about 4.8 million actions) we directly used the keyword and category information. In particular, for a *view_record* action we used the top three keywords and the main classification of the viewed document. For a *search* action we computed the most assigned keywords and classification from the resulting documents and used the top three keywords and the most common classification.

---

[1] http://sowiport.gesis.org

[2] http://www.gesis.org/en/services/research/tools-zur-recherche/social-science-thesaurus/

[3] http://www.gesis.org.org/unser-angebot/recherchieren/tools-zur-recherche/klassifikation-sozialwissenschaften/

For user actions without SOLIS documents (about 3.8 million actions) we applied a number of processing steps. If the record(s) contained keyword information we transformed those to equivalent SOLIS keywords with the help of the HTS service. If not, we used the title of the record and user search terms, cleaned them from stop words and other specialized characters and transformed them to SOLIS keywords with the help of the STR service. If the action then has SOLIS keyword information we applied the top category learned from the existing actions for the top three keywords.

We are aware that the application of these methods gives only an abstraction of the users' true information needs. The quality of capturing them thereby depends on the quality of the document collection, on the ranking functionality and the quality of the HTS and STR services. However, the application of these methods allowed large scale clustering of search interests and we were able to add keywords and categories to 8,605,189 user actions.

## 5. VISUALIZATIONS

The main question for the visualization part is how we can achieve a clear and meaningful presentation of more than eight million user actions including categorization, keywords and user search terms. The main focus here is to get an instant overview of user interests at each level of detail. We have therefore created an interactive visualization which allows an easy analysis of specific sub fields without getting lost in the data. To this aim we exploited the hierarchy of the classification system and decided to arrange the user actions on four zoomable levels: (1) main categories, (2) subcategories, (3) keywords, and (4) user search terms. A path in this hierarchy can be for example: Sociology → Sociology of Economics → Income → Social Inequality. The prototype can be found at http://vizgr.org/social-science-topics.

The first level gives an overview of the 14 main categories from the classification for the social sciences. For the visualization we used the circle pack layout[4] of the D3.js framework. In this visualization type the circle radius of the main categories (level one) and the subcategories (level two) is built recursively by keywords with more than 500 user actions on the third level (this amount is still doable by the framework). This way, the visualization is an abstraction of all user actions and shows only the most interesting categories and topics users are interested in. Exact proportions given in the tables of this paper are computed from the database table. In Figure 1, we can get an instant overview on which top categories were most frequently accessed by users. For example, the third largest main category is "Interdisciplinary and Applied Field of the Social Sciences".

With a mouse click on a category we can zoom in. Now, all categories within the top category are shown. 145 sub classes are ordered in the top categories. Here, the circle radius encodes the number of actions within this category. This way we can get an instant overview on which categories were most frequently accessed by users. For example, in Figure 2, we can see the subcategories of "Interdisciplinary and Applied Field of the Social Sciences".

With a second mouse click on a category bubble we can zoom into that certain category. Now, all keywords with more than 500 user actions are shown (Figure 3 shows the topics for the category "Social Problems"). Again, the circle radius encodes the number of user actions with this keyword. We can see which topics appear most frequently in this specific category.

With a third mouse click on a keyword bubble we can zoom into that keyword. Now we see most frequent user search terms combinations for this keyword (here for the topic "Discrimination" shown in Figure 4). For this we used a tree layout[5]. The tree layout is separated by several levels from left to right. On the first level (tree node) we see the document keyword. On the second level it is shown which user search terms were frequently used for this document keyword. The strength of relationship is visualized by the ordering (from top to bottom) and additionally by different edge width and the node radius. On the second level we can then see which second user term has been combined with the prior term (and so on). This way, we can see instantly which user search term combinations appear most frequently for a given document keyword.

## 6. ANALYSIS

Having applied the same classification system to all user actions, we now have a good overview on which main categories, subcategories and topics users are interested in. The visualization gives us the chance to dive into the data and to have an instant overview on what is interesting.

### 6.1 Main Categories

The clustering by main categories gives us an overview on an abstract level which broad fields of the social sciences our users are interested in. Table 1 shows the main categories ordered by the percentage of user actions. The top 5 main categories are (M1) Sociology, (M2) Interdisciplinary and Applied Fields of the Social Sciences, (M3) Political Science, (M4) Education and Pedagogics and (M5) Science of Communication. The percentage of user actions shows that a good portion of about 73% of the users is interested in these five fields. Of these, Sociology is by far the largest category with about 28%, followed by Interdisciplinary and Applied Fields of the Social Sciences with 15% and Political Science with 13%.

We find it hard to identify subject distributions from prior studies on the social science domain that we could compare our findings to. In fact, even the term "social sciences" is not defined consistently in the literature. In 1983, Haart [7] gives an overview of the different definitions of social sciences in different countries and classification systems (Dewey, UDC, UNESCO). For Germany, the selection of the main fields of the social sciences in his and our table is nearly identical. Some evidence about the distribution is given from the field of citation analysis. Early work from Earle & Vickery [4] in 1965 found Politics, Economics, Education and Sociology to be the largest subjects by citation count in social science literature of the UK. Line [17] in 1981 found Psychology, Sociology and economics to be predominant from citation counts of 300 monographs and 140 serials. Haart in 1983 [7] reports from the DISISS project with 2,760 serials that the subjects Economics, Education, Political Science and Psychology were prominent. We were unable to find more recent material showing the distribution of social science subjects by citation analysis with an adequate distinction of subject fields.

---

[4] https://github.com/mbostock/d3/wiki/Pack-Layout

[5] https://github.com/mbostock/d3/wiki/Tree-Layout

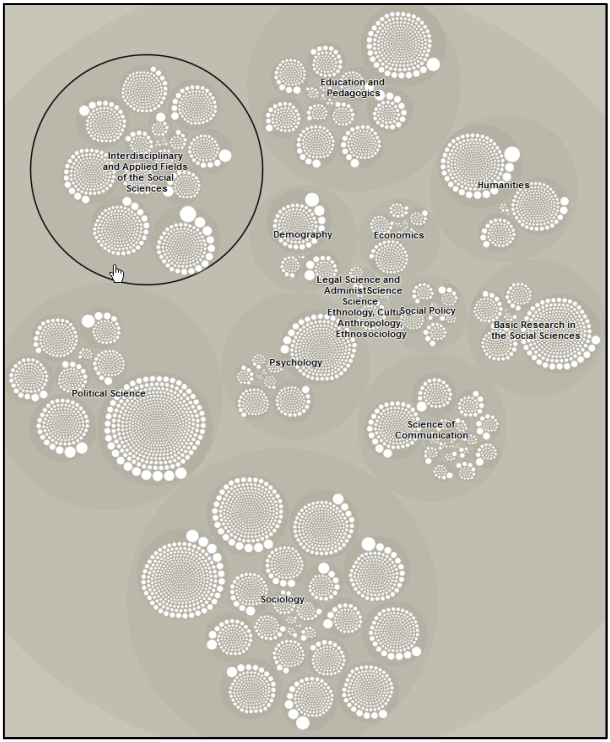

**Fig. 1. First level: Overview of the main categories**

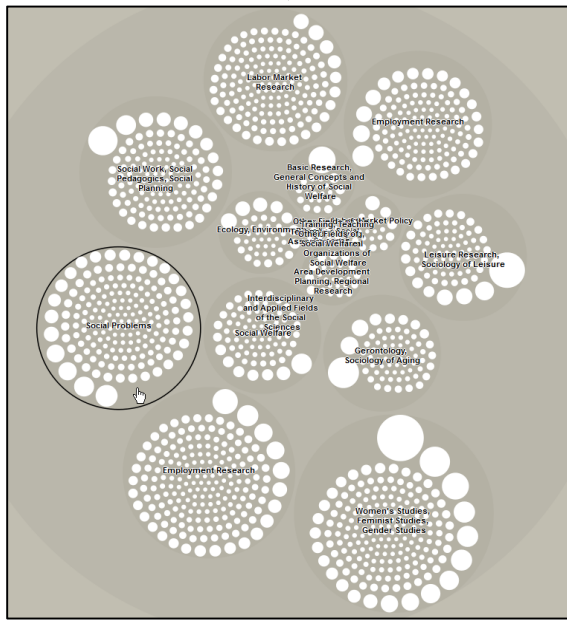

**Fig. 2. Second level: Overview of the subcategories in "Interdisciplinary and Applied Field of the Social Sciences"**

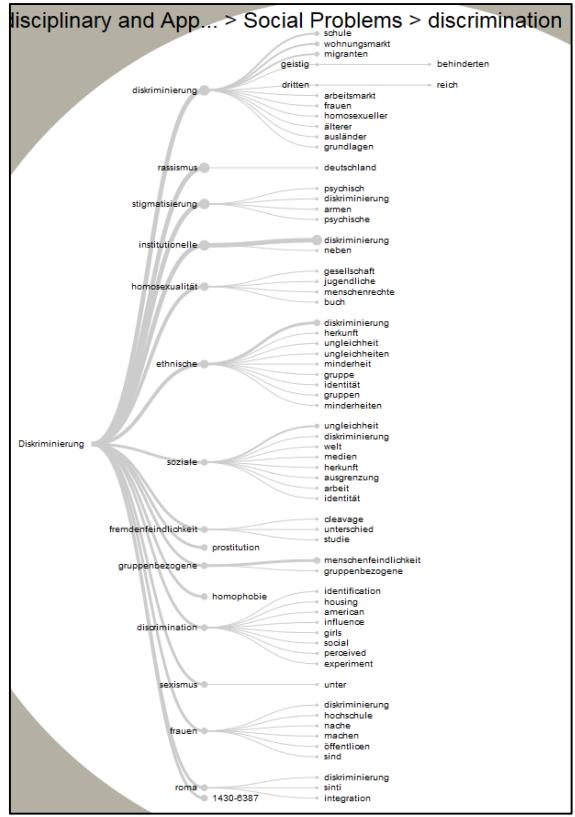

**Fig. 4. Fourth level: User search terms for the topic "Discrimination". Not translated as these are user terms.**

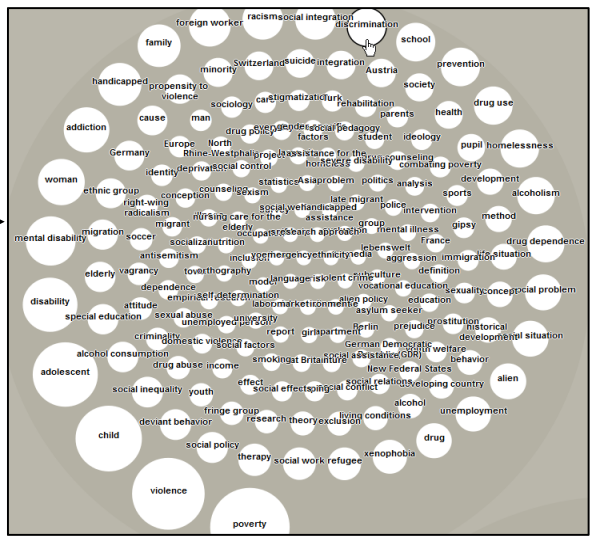

**Fig. 3. Third level: Topics in the category "Social Problems"**

## 6.2 Subcategories

On the second level we want to have a look at the subcategories users are interested in. Table 2 gives an overview of the top twenty subcategories. The top five subcategories are (S1) Political Process, Elections, Political Sociology, (S2) General Sociology, (S3) Social History, (S4) Methods and Techniques of Data Collection and Data Analysis and (S5) Sociology of Education. These five subcategories make up about 20% of all user actions.

Table 1. Main categories ordered by user interest.

| # | Main category | Percent of User Actions |
|---|---|---|
| M1 | Sociology | 28.00% |
| M2 | Interdisciplinary and Applied Fields of the Social Sciences | 14.99% |
| M3 | Political Science | 13.27% |
| M4 | Education and Pedagogics | 10.53% |
| M5 | Science of Communication | 5.99% |
| M6 | Humanities | 5.98% |
| M7 | Psychology | 5.88% |
| M8 | Basic Research in the Social Sciences | 5.61% |
| M9 | Demography | 2.99% |
| M10 | Economics | 2.78% |
| M11 | Social Policy | 2.28% |
| M12 | Ethnology, Cultural Anthropology, Ethnosociology | 0.70% |
| M13 | Legal Science and Administrative Science | 0.68% |
| M14 | Science, Engineering, Medicine | 0.29% |

The visualization in Figure 1 allows us some interesting observations: (I) It instantly clarifies that top subcategories are distributed over the top nine main categories (we can see that there is at least one big bubble in the first nine main categories). This lets us conclude that any user's interest is usually well distributed over the main classes. (II) The main category Sociology contains a lot of specialized sociologies. Here, the structures and processes of thematic sub-areas or social groups are examined. Eight subcategories are included in the top 20 (see Table 2) with all different aspects of Sociology such as Education, Culture, Art, Family, Industry and Work, Youth, Criminality, Medicine, Communication, Religion and so on. However, also all other specialized sociologies are frequently questioned by users (compare Figure 1 to see the number and size of bubbles included in Sociology) which in sum makes Sociology the most prominent main category. (III) The main category Political Science contains two important subcategories: (a) Political Process, Elections, Political Sociology and (b) Peace and Conflict Research, International Conflicts. (IV) The main category "Interdisciplinary and Applied Fields of the Social Sciences" is represented by three important subcategories from Table 2: (a) Woman's and Gender Studies, (b) Employment Research and (c) Social Problems. (V) Beyond the classification by main categories, the topic Education plays an important role with the two subcategories "Sociology of Education" (S5) from M1 and "Education and Pedagogics" (S7) from M4 forming together 6.07 % of all user actions.

To follow the distribution of subcategories over time, we built another visualization. Fig. 5 shows the distribution of categories from 2007 to 2016 for the top ten categories in percent of user actions per month in this category. It can be seen that the overall trend for most categories seems to be relatively constant with variations of a few percentage points. We examined this for the top 30 categories and found that there is a solid and constant level of interest in the categories. Exceptions are the subcategories that deal with the topic Education. Here, we can see a growing interest over time, especially in the beginning of 2014 when the Sowiport portal was relaunched.

Table 2. Top 20 subcategories users are interest in.

| # | Subcategory | Main category | Percent of user actions |
|---|---|---|---|
| S1 | Political Process, Elections, Political Sociology | M3 | 5.98% |
| S2 | General Sociology, Basic Research, General Concepts and History | M1 | 4.02% |
| S3 | Social History | M5 | 3.30% |
| S4 | Methods and Techniques of Data Collection and Data Analysis | M7 | 3.23% |
| S5 | Sociology of Education | M1 | 3.20% |
| S6 | Social Psychology | M8 | 2.93% |
| S7 | Education and Pedagogics | M4 | 2.87% |
| S8 | Women's Studies, Feminist Studies, Gender Studies | M2 | 2.36% |
| S9 | Cultural Sociology, Sociology of Art, Sociology of Literature | M1 | 2.28% |
| S10 | Family Sociology, Sociology of Sexual Behavior | M1 | 2.07% |
| S11 | Employment Research | M2 | 2.04% |
| S12 | Peace and Conflict Research, International Conflicts | M3 | 2.02% |
| S13 | Sociology of Work, Industrial Sociology, Industrial Relations | M1 | 1.90% |
| S14 | Migration, Sociology of Migration | M9 | 1.88% |
| S15 | Science of Communication | M6 | 1.81% |
| S16 | Social Problems | M2 | 1.78% |
| S17 | Criminal Sociology, Sociology of Law | M1 | 1.73% |
| S18 | Medical Sociology | M1 | 1.72% |
| S19 | Philosophy, Religion | M5 | 1.68% |
| S20 | Sociology of Youth, Sociology of Childhood | M1 | 1.57% |

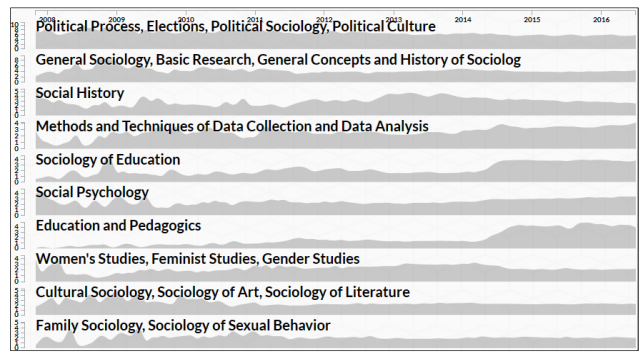

Fig. 5. Top ten categories from 2007 to 2016

## 6.3 Topics in Categories

On the third level we can explore which topics appear in a category with at least 500 matching user actions. Subcategories with many diverse topics may encounter a lot of aspects of the topic (e.g. for S2 General Sociology: Social theory, Society, Social Change etc.). Additionally many geographic aspects can appear (e.g. for S1 Political Process, Elections: Great Britain, France, Developing Country etc.). Also, temporal aspects can be found (e.g. in S2 Social History: 20$^{th}$ century, 19$^{th}$ century etc.). This level spans thus a network with all different aspects that touch the subject. This view on categories is new as it is not explicitly discarded in the category system but implicitly built by keywords of viewed and searched documents in this category and chosen by user interaction. Table 3 shows the main topics for the five most viewed subcategories.

**Table 3. Main topics for subcategories.**

| # | Subcategory | Top five topics |
|---|---|---|
| S1 | Political Process | democracy, party, politics, Great Britain, right-wing radicalism |
| S2 | General Sociology | sociology, sociological theory, theory, society, social inequality |
| S3 | Social History | teaching materials, mathematics instruction, vocational education, school, sports |
| S4 | Data Collection and Analysis | qualitative method, interview, measurement, empirical social research, research |
| S5 | Sociology of Education | school, education, teaching, university, Germany |

A more detailed example is shown in Figure 3 with the topics for the subcategory S16 "Social Problems". Here, about 111 different aspects of social problems are visualized. From generic problems such as poverty, violence, disability, addiction, alcoholism, racism to social units such as child, youth, woman, parents to different social groups such as foreign employee, foreigners, elder people, disabled peoples, refugees, Turks and countries such as Suisse, Austria, France to provisions such as rehabilitation, prevention, social integration, therapy and so on.

## 6.4 User terms for a topic

On the most specific level we leave the predefined classification system of categories and keywords curated by information professionals and enter the world of user search terms. This level explicitly indicates in which aspects users are interested by presenting their query terms in combination. We computed the count of user terms for all search actions and found that on average users enter 2.21 terms per query. This is similar to the average count of terms found in web search by Jansen et al. [12] with 2.21 terms. For domain-specific search different values reported: for Health Records: 1.2 terms [23], for E-journals in medicine: 2.27 terms [14] and for Pubmed: 3.54 terms [10]. For Psychology: 3.16 terms and History 3.42 terms [30] and for the field of Computer Science: 2.43 terms [13] and "three or fewer" [20].

As an example for this level of user terms we use Figure 4. It shows the word tree for the topic "discrimination". On the first level important user terms are discrimination, racism, stigmatization, institutional, xenophobia, prostitution, ethnic, social, sexism, homophobia, homosexuality, women, group-focused, sexual, antiziganism, prejudice and Roma. On the second level these terms combine for example to "discrimination school", "discrimination migrants", "institutional discrimination" or "group-focused enmity".

## 7. INDEXING LANGUAGE VS. USER LANGUAGE

Another aspect we want to address here is what differences exist between keywords applied to documents by information professionals (level 3 in our visualization) and the terms users apply for searching (on level 4). This issue has been addressed in research as the "vocabulary problem" [6]. Users use a different vocabulary to articulate their information needs than the vocabulary used to index the documents in a collection.

To investigate this issue in detail it is helpful to understand the way how information professionals assign index keywords to documents. As already described the thesaurus for the social sciences contains about 12,000 entries with 8,000 descriptors and 4,000 synonyms (non-descriptors) which refer to these descriptors. All entries are linked with semantic relationships such as "related", "broader", "narrower", "use instead", "used for" and are enriched with notes for disambiguation and a classification assignment. Similar to the constant development of a scientific discipline, also a thesaurus is in constant further development. So, new descriptors and non-descriptors are included in the thesaurus after carefully weighing up their importance. The vocabulary of a thesaurus should be as small as possible to make the indexing process consistent and retrieval results precise, but also as large as necessary to capture all phenomena of a scientific discipline. One method is to set each descriptor in nominative singular to avoid all declinations of a term. A second methodology to keep the vocabulary relatively small is the principle of semantic decomposition. Compound terms such as "bittersweet" are decomposed into individual terms like "bitter" and "sweet" as long as no significant loss of meaning is accompanied. These copulative compounds are very frequently used in the German language. As a result the content of documents is described with several equally weighted descriptors (post-coordinate indexing [26]).

There are several approaches to mitigate the problem of different languages of users and the system, for example with term recommenders. In Sowiport we have included such a recommender which helps the user in the process of formulating the search query. This is done with two methods: first, by autocompleting user entries with thesaurus terms and second, by showing descriptors that are semantically near [9]. In a long term evaluation we could show that the term recommender in Sowiport is used in 10% of the cases and has a largely positive effect for the following search processes [8].

However, users still use their own language for several reasons. First and foremost, users might not be aware that using a controlled vocabulary leads to search results with higher precision and recall. Here, the difference between web search engines and domain specific digital libraries becomes visible. Web search engines nearly always return results due to the immense corpus of web pages. In domain specific digital libraries the usage of knowledge organization systems is still doable and leads to better results if the vocabulary is used. A second reason might be that the users have information needs that cannot be expressed with the existing vocabulary. Be it new phenomena in a discipline that still cannot be represented with thesaurus terms or are circulated under several different terms, or phenomena in detail that are under the level of existing (combinations of) descriptors.

## 7.1 Comparing user search queries to thesaurus terms

In the following we will measure the overlap and especially the difference between vocabulary used on the system and on the user side. This can in particular show if information needs of real users can be expressed with the existing thesaurus and on the other side which needs cannot be expressed and are possible candidates for the inclusion in the thesaurus.

**Table 4. Example user search terms not found in the thesaurus.**

| # | Original Term | English Translation | Example Search Query | Thesaurus | Category |
|---|---|---|---|---|---|
| 1 | tokyo | tokyo | city development tokyo | Could be included, used in more than 5,930 docs | Basic |
| 2 | freiwilligenmanagement | voluntary management | voluntary managment elderly | Can be described by separate descriptors | Remain current |
| 3 | jugendgewalt | youth violence | youth violence xenophobia | Can be described by separate descriptors | Basic |
| 4 | seniorenarbeit | work with elderly | community oriented work with elderly | Can be described by separate descriptors | Basic |
| 5 | pegida | pegida (far-right movement founded in Dresden in October 2014 which promotes anti-islamic positions) | refugee pegida | Could be included, used in more than 30 docs | Lately |
| 6 | pflegende | nourishing | nourishing relatives | Exists in nominative singular | Basic |
| 7 | kita | abbreviation for "Kindertagesstätte" - day nursery | day nursery studies | Could be included as non-descriptor | Remain current |
| 8 | dating | dating | online dating | The German (broader) term "Partnersuche" could be included and "dating" as a synonym | Remain current |
| 9 | kontakthypothese | contact hypothesis | contact hypothesis xenophobia | Could be included, used in more than 49 docs | Specialized |
| 10 | ttip | ttip | ttip public opinion | Could be a narrower term for "Freihandel" | Lately |
| 11 | geschlechterrollen | gender roles | children gender roles | Could be included as non-descriptor | Basic |
| 12 | blame | blame | culture of blame | Could be included as translation | Basic |
| 13 | eurokrise | euro crisis | solidarity euro crisis | Could be included, used in more than 280 docs | Remain current |
| 14 | praise | praise | praise and blame | Could be included, used in more than 2,728 docs | Basic |
| 15 | sharing | sharing | sharing economy | Could be included, used in more than 24,409 docs | Remain current |
| 16 | migrantinnen | migrant woman | older migrant woman | Can be described by separate descriptors | Basic |
| 17 | internetnutzung | internet use | internet use youth | Can be described by separate descriptors | Basic |
| 18 | herkunftseffekte | social origin factors | primary and secondary effects of social origin factors | Could be included, used in more than 86 docs | Specialized |
| 19 | zukunftsplanung | future planning | personal future planning | Could be included, used in more than 280 docs | Specialized |
| 20 | ageing | ageing | ageing population | "Ageing society" could be included | Basic |
| 21 | willkommenskultur | welcome culture | refugees welcome culture | Could be included, used in more than 64 docs | Lately |
| 22 | bildungsvererbung | education inheritance | Education inheritance and health | Could be included, used in more than 30 docs | Specialized |
| 23 | lebensweltorientierung | lifeworld orientation | social work and lifeworld orientation | Could be included, used in more than 200 docs | Specialized |
| 24 | biografiearbeit | biography work | biography work dementia | Could be included, used in more than 200 docs | Basic |
| 25 | heimatgefühl | sense of home | sense of home and identity | Could be included, used in more than 70 docs | Lately |

For the analysis we used the same dataset as in the previous chapter. In particular we used all search queries from users that had been entered into a search form (simple or advanced) with the field type set to "all" (default) or "keyword" with about 1.7 million search queries. We found that users often copy ISSNs of journals or enter person names into the search bar. We filtered these queries as first approach with regular expressions. We then tokenized each query into individual tokens, applied a German and English stop word list, filtered special characters and finally used a German Porter stemmer to reduce declinations to their word stem. We found that there were still a large number of person names that are hard to capture by regular patterns. Therefore, we compared tokens to all author names from Sowiport (about 35,000 distinct first and last names) and, if found, erased them from the test set. Finally, each token was compared to all German and English translations in the thesaurus to find if it was included or not. For all terms not found in the thesaurus we also counted the frequency.

As a final result we found that about 88.26% of the user terms could be found in the thesaurus. That is really a large part, which means that most information needs could be expressed with the existing vocabulary of the thesaurus.

However, much more insight can be given by the part which could not be found. After all filtering steps 11.74% of the terms in user search queries could not be found in the thesaurus (143,934 of 1,226,503 terms; 40,965 distinct terms). Figure 6 shows that the frequency distribution of the missing terms follows a really long tail. That means that only a few new terms have been entered by users regularly, but there is a really long tail of terms which has been entered only once or twice. Even in this filtered list, we still have a lot of terms which have no specific topical meaning ("approach", "based", "way", "related", "versus", "survey", "handbook") and which make only sense when used with a second or third term.

Table 4 shows some examples of frequent terms which have not been found in the thesaurus. Among them are very recent social phenomena or movements such as "Pegida" (since 2014), "TTIP" (since 2014), "welcome culture" (2015) or "sense of home" (2015). "Pegida" is a far-right movement founded in Dresden in October 2014 which promotes regularly anti-islamic demonstrations. "TTIP" is a trade agreement between the US and the European Union under ongoing negotiations since 2013 and which is discussed controversially. "Welcome culture" in this context means the opening of the German society towards refuges within the European migrant crisis in 2015. "Sense for home" is a reoccurring phenomenon in Germany where sense for home is enforced by the globalized and highly industrialized world. Characteristic is that all of these terms show only small numbers of documents in the collection.

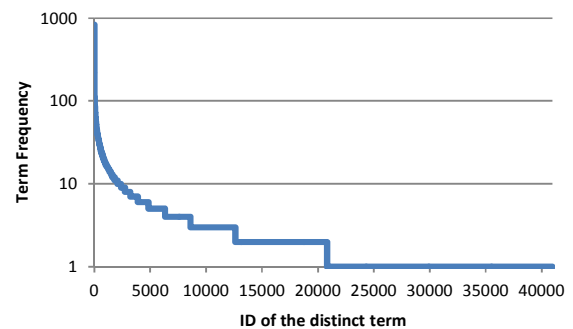

**Fig. 6: Frequency of user search terms which have not been found in the thesaurus**

Furthermore, there are several topics and phenomena which already run over some years, but also started lately in our logs. For example the "euro crisis" has started in 2010. The topic of "sharing" has started with "file sharing" in 2009 and then developed over "car sharing", "job sharing", "music sharing", "bike sharing" and "house sharing" until 2016. The topic "day nursery" started in 2010. Day nursery is a recent topic in Germany, especially for children from one to three years old. In 2013 a law went into effect that states a legal claim for a nursery place for this group of children. For the next topic "voluntary management" there is much more experience and literature in the Anglo-American World and it has come up in Germany lately and in our logs since 2012.

Then we can compare between basic concepts on the one side and specialized concepts on the other. "Tokyo" seems to be requested very often and can be found regularly in the literature. It seems that the thesaurus has not focused on that yet. "Biography work" is an essential method in the social sciences and could also be included. The topics "pflegende (nourishing)" and "Migrantinnen (migrant woman)" show the limits of the principle of semantic decomposition for indexing in the German language: "Migrant UND Frau" is not the same as "Migrantinnen", as the first describes "migrant AND woman" and the second "female migrants". The same applies for the second example: "pflegende Eltern" is not the same as "Pflege UND Eltern" as the first describes "parents who are nourishing" and the second describes "care and parents". Specialized terms such as "kontakthypothese" ("contact hypothesis"), "herkunftseffekt" ("social origin factors"), "zukunftsplanung" ("future planning"), "bildungsvererbung" ("education inheritance") or "lebensweltorientierung" ("lifeworld orientation") have a similar challenge in the German language (as opposed to the English language) as the semantic decomposition of these terms do not describe the same concept as the compound term.

However, as we can see the comparison between the system- and the user language can show important new and recent topics and aspects of user interest.

## 8. CONCLUSION

To answer *research question (1)* "What are the user interests in the (German) Social Sciences on different topical levels and over a long period of time?" we have analyzed nine years transaction logs from a digital library for the social sciences to study the topical interests of German users in this field. On the level of main categories there is a strong interest for (1) Sociology, (2) Interdisciplinary and Applied Fields of the Social Sciences, (3) Political Science, (4) Education and Pedagogics and (5) Science of Communication. Sociology is by far the most important category with about 28% of interest. The analysis on the level of subcategories show a strong interest in (a) Political Process, Elections, Political Sociology, (b) General Sociology, (c) Social History, (d) Methods and Techniques of Data Collection and Data Analysis and (e) Sociology of Education. The top subcategories are evenly distributed over the main categories with a strong focus on Sociology. This means on the one hand that the user's interest is well distributed over all main categories and on the other that there is a strong focus on Sociology and its sub disciplines. We could see that on the level of subcategories the user interests are stable over time. On the next more specific level, the level of topics, we can see the diversity of interest within a subcategory chosen by user interaction. On the fourth level we then finally show user search terms and here we can see which user search terms were used in combination. All in all, the categorization of user actions into the classification system is a good method to handle the topical complexity of the field of social sciences.

*Research question (2)* asked "Are knowledge organization systems such as thesauri good enough to capture these user interests?". We therefore compared in detail the indexing language on level 3 and the user language on level 4. We found that a large part of over 88% of user search terms can be found in the thesaurus, which shows the power and topical coverage of the thesaurus system. Thus, the thesaurus seems to be an important instrument to cover a majority of user interests but at the same time to allow precise retrieval. On the other hand, the 12% of the topics which could not be found in the thesaurus show interesting trends of longer lasting and very recent new research questions. These user topics are good candidates for the inclusion into the thesaurus. Comparing indexing and user language seems to be a good method to identify actual strengths and weaknesses of a thesaurus. This allows among others to find topics and themes which are no longer interesting (or have never been) and can therefore be deleted.

## ACKNOWLEDGMENTS

This work was partly funded by the DFG, grant no. MA 3964/5-1; the AMUR project at GESIS. The author thanks the focus group IIR at GESIS for fruitful discussions and suggestions.

## 9. REFERENCES

1. Maristella Agosti, Franco Crivellari, and Giorgio Maria Di Nunzio. 2011. Web log analysis: a review of a decade of studies about information acquisition, inspection and interpretation of user interaction. *Data Mining and Knowledge Discovery* 24, 3: 663–696. https://doi.org/10.1007/s10618-011-0228-8
2. Philip M. Davis. 2004. Information-seeking behavior of chemists: A transaction log analysis of referral URLs. *JASIST* 55, 4: 326–332. https://doi.org/10.1002/asi.10384
3. Philip M. Davis and Leah R. Solla. 2003. An IP-level analysis of usage statistics for electronic journals in chemistry: Making inferences about user behavior. *Journal of the American Society for Information Science and Technology* 54, 11: 1062–1068. https://doi.org/10.1002/asi.10302
4. Penelope Earle and Brian Vickery. 1969. Social science literature use in the UK as indicated by citations. *Journal of Documentation* 25, 2: 123–141. https://doi.org/10.1108/eb026468
5. D. Ellis, D. Cox, and K. Hall. 1993. A comparison of the information seeking patterns of researchers in the physical and social sciences. *Journal of Documentation* 49: 356–356.

6. G. W. Furnas, T. K. Landauer, L. M. Gomez, and S. T. Dumais. 1987. The Vocabulary Problem in Human-system Communication. *Commun. ACM* 30, 11: 964–971. https://doi.org/10.1145/32206.32212
7. H. P. Hogeweg-De Haart. 1983. Characteristics of social science information: A selective review of the literature. Part I. *Social Science Information Studies* 3, 3: 147–164. https://doi.org/http://dx.doi.org/10.1016/0143-6236(83)90021-2
8. Daniel Hienert and Peter Mutschke. 2016. A Usefulness-based Approach for Measuring the Local and Global Effect of IIR Services. In *Proceedings of the 2016 ACM on Conference on Human Information Interaction and Retrieval* (CHIIR '16), 153–162. https://doi.org/10.1145/2854946.2854962
9. Daniel Hienert, Frank Sawitzki, and Philipp Mayr. 2015. Digital Library Research in Action: Supporting Information Retrieval in Sowiport. *D-Lib Magazine* 21, 3/4. https://doi.org/10.1045/march2015-hienert
10. Rezarta Islamaj Dogan, G. Craig Murray, Aurelie Neveol, and Zhiyong Lu. 2009. Understanding PubMed user search behavior through log analysis. *Database : the journal of biological databases and curation* 2009: bap018. https://doi.org/10.1093/database/bap018
11. Bernard J. Jansen. 2006. Search log analysis: What it is, what's been done, how to do it. *Library & Information Science Research* 28, 3: 407–432. https://doi.org/http://dx.doi.org/10.1016/j.lisr.2006.06.005
12. Bernard J. Jansen, Amanda Spink, and Tefko Saracevic. 2000. Real Life, Real Users, and Real Needs: A Study and Analysis of User Queries on the Web. *Inf. Process. Manage.* 36, 2: 207–227. https://doi.org/10.1016/S0306-4573(99)00056-4
13. Steve Jones, Jo Sally Cunningham, Rodger McNab, and Stefan Boddie. A transaction log analysis of a digital library. *International Journal on Digital Libraries* 3, 2: 152–169. https://doi.org/10.1007/s007999900022
14. Hao-Ren Ke, Rolf Kwakkelaar, Yu-Min Tai, and Li-Chun Chen. 2002. Exploring behavior of E-journal users in science and technology: Transaction log analysis of Elsevier's ScienceDirect OnSite in Taiwan. *Library & Information Science Research* 24, 3: 265–291. https://doi.org/http://dx.doi.org/10.1016/S0740-8188(02)00126-3
15. Diane Kelly. 2009. Methods for Evaluating Interactive Information Retrieval Systems with Users. *Found. Trends Inf. Retr.* 3, 1—2: 1–224. https://doi.org/10.1561/1500000012
16. Diane Kelly and Colleen Cool. 2002. The Effects of Topic Familiarity on Information Search Behavior. In *Proceedings of the 2Nd ACM/IEEE-CS Joint Conference on Digital Libraries* (JCDL '02), 74–75. https://doi.org/10.1145/544220.544232
17. Maurice B. Line. 1981. The structure of social science literature as shown by a large-scale citation analysis. *Social Science Information Studies* 1, 2: 67–87. https://doi.org/http://dx.doi.org/10.1016/0143-6236(81)90001-6
18. Maurice B Line. 2000. Social Science Information-The Poor Relation. *IFLA Journal-International Federation of Library Associations* 26, 3: 177–179.
19. Thomas Lüke, Philipp Schaer, and Philipp Mayr. 2012. Improving Retrieval Results with Discipline-Specific Query Expansion. In *TPDL* (Lecture Notes in Computer Science), 408–413. Retrieved from http://arxiv.org/abs/1206.2126
20. Malika Mahoui and Sally Jo Cunningham. 2001. Search Behavior in a Research-Oriented Digital Library. In *Research and Advanced Technology for Digital Libraries: 5th European Conference, ECDL 2001 Darmstadt, Germany, September 4-9, 2001 Proceedings*, Panos Constantopoulos and Ingeborg T. Sølvberg (eds.). Springer Berlin Heidelberg, Berlin, Heidelberg, 13–24. Retrieved from http://dx.doi.org/10.1007/3-540-44796-2_2
21. Philipp Mayr and Vivien Petras. 2008. Cross-concordances - terminology mapping and its effectiveness for Information retrieval: Crosskonkordanzen - Terminologie Mapping und deren Effektivität für das Information Retrieval. In *World Library and Information Congress*. Retrieved from http://archive.ifla.org/IV/ifla74/papers/129-Mayr_Petras-en.pdf
22. Lokman I. Meho and Helen R. Tibbo. 2003. Modeling the Information-seeking Behavior of Social Scientists: Ellis's Study Revisited. *J. Am. Soc. Inf. Sci. Technol.* 54, 6: 570–587. https://doi.org/10.1002/asi.10244
23. Karthik Natarajan, Daniel Stein, Samat Jain, and Noémie Elhadad. 2010. An Analysis of Clinical Queries in an Electronic Health Record Search Utility. *International journal of medical informatics* 79, 7: 515–522. https://doi.org/10.1016/j.ijmedinf.2010.03.004
24. David Nicholas, Paul Huntington, and Janet Homewood. 2003. Assessing used content across five digital health information services using transaction log files. *Journal of Information Science* 29, 6: 499–515. https://doi.org/10.1177/0165551503296007
25. Thomas A. Peters. 1993. The history and development of transaction log analysis. *Library Hi Tech* 11, 2: 41–66. https://doi.org/10.1108/eb047884
26. M. Taube and Documentation Incorporated. 1953. *Studies in Coordinate Indexing*. Documentation Incorporated.
27. Luis Villen-Rueda, Jose A. Senso, and Felix Moya-Anegon. 2007. The Use of OPAC in a Large Academic Library: a transactional log analysis of subject searching. *The Journal of the Academic Librarianship* 33: 327–337. https://doi.org/10.1016/j.acalib.2007.01.018
28. Ryen W. White, Susan T. Dumais, and Jaime Teevan. 2009. Characterizing the Influence of Domain Expertise on Web Search Behavior. In *Proceedings of the Second ACM International Conference on Web Search and Data Mining* (WSDM '09), 132–141. https://doi.org/10.1145/1498759.1498819
29. Barbara M. Wildemuth. 2004. The Effects of Domain Knowledge on Search Tactic Formulation. *J. Am. Soc. Inf. Sci. Technol.* 55, 3: 246–258. https://doi.org/10.1002/asi.10367
30. Kwan Yi, Jamshid Beheshti, Charles Cole, John E. Leide, and Andrew Large. 2006. User search behavior of domain-specific information retrieval systems: An analysis of the query logs from PsycINFO and ABC-Clio's Historical Abstracts/America: History and Life. *Journal of the American Society for Information Science and Technology* 57, 9: 1208–1220. https://doi.org/10.1002/asi.20401